\documentclass[aps,prb,reprint,showpacs,superscriptaddress,groupedaddress]{revtex4-1}
\usepackage{graphicx}
\usepackage{amsmath}
\usepackage{amssymb}
\usepackage{dcolumn}
\usepackage{dsfont}
\usepackage{latexsym}
\usepackage{rotating}
\usepackage{bbm}
\usepackage[usenames,dvipsnames]{xcolor}  
\usepackage{float}
\usepackage{epsfig}
\usepackage{psfrag}
\usepackage{natbib}
\usepackage{bm}
\usepackage{eucal}
\usepackage{mathrsfs}
\usepackage{braket}
\usepackage{enumerate}
\usepackage{verbatim}
\usepackage{longtable}
\usepackage{subfigure}
\usepackage{bm}
\usepackage[breaklinks,hyperindex]{hyperref}
\usepackage{amsfonts}
\usepackage{dcolumn}
\usepackage{bm}
\usepackage{subfigure}

\newcommand{\be}{\begin{equation}}
\newcommand{\ee}{\end{equation}}
\newcommand{\bn}{\begin{eqnarray}}
\def\dc{{\text{dc}}}
\def\hc{\text{h. c.}}
\def\HFKM{\text{HFKM}}
\def\FL{\text{FL}}

\usepackage{color} 

\usepackage{hyperref}
\hypersetup{
colorlinks=true,final=true,
        linkcolor=red,
        citecolor=blue,
        filecolor=blue,
        urlcolor=blue,
}
\begin{document}

\title{ Realization of a ``Two Relaxation Rates'' in the Hubbard-Falicov-Kimball Model}

\author{H. Barman$^{1}$}\email{hbarhbar@gmail.com}
\author{M. S. Laad$^{1}$}\email{mslaad@imsc.res.in}
\author{S. R. Hassan$^{1}$}\email{shassan@imsc.res.in}

\affiliation{$^{1}$Institute of Mathematical Sciences, Taramani, Chennai 600113, India
}

\begin{abstract}
  A single transport relaxation rate governs the decay of both, longitudinal and Hall currents in Landau Fermi Liquids
(LFL).  Breakdown of this fundamental feature, first observed in cuprates and subsequently in other 
{\it three-dimensional} correlated systems close to (partial or complete) Mott metal-insulator transitions, played a 
pivotal role in emergence of a non-Landau Fermi liquid paradigm in higher dimensions $D(>1)$.  Motivated hereby, we 
explore the emergence of this ``two relaxation rates'' scenario in the Hubbard-Falicov-Kimball model (HFKM) using 
the dynamical mean-field theory (DMFT).  Specializing to $D=3$, 
we find, beyond a critical FK interaction, that two distinct relaxation rates governing distinct temperature 
($T$) dependence of the longitudinal and Hall currents naturally emerges in the non-LFL metal.  We rationalize this 
surprising finding by an analytical analysis of the structure of charge and spin correlations in the underlying 
impurity problem, and point out good accord with observations in the famed case of V$_{2-y}$O$_3$ near the MIT. 
\end{abstract}



\maketitle

  It is well known that a single transport relaxation rate governs the decay of both longitudinal and Hall currents in a Landau Fermi liquid (LFL) metal.  This is obviously related to the fact that both result from scattering processes
involving the same Landau quasiparticle, carrying the quantum numbers of an electron.  Observation of distinct relaxation rates in dc resistivity ($\rho_{\text{dc}}$) and Hall angle ($\theta_H$)~\cite{chien1991} data for cuprates led to a paradigm shift
in the traditional view of strongly correlated electrons in metals in dimension $D>1$.  While such anomalous behavior 
can be rationalized in $D=1$ Luttinger liquids (LL) by appealing to fractionalization of an electron into a neutral 
spinon and a spinless holon, the specific nature of electronic processes leading to emergence of such features in $D>1$ is an enigma.  In fact, Anderson~\cite{anderson1991} {\it predicted} such a feature from a generalized ``tomographic'' LL
state in $D=2$, by hypothesizing spin-charge separation: $\rho_{\text{dc}}(T)\simeq T$ arose from holon-spinon scattering, 
while cot$\theta_{\text{H}}(T)\simeq T^{2}$ emerged from spinon-spinon scattering.  Though very attractive, a derivation of
such a higher-$D=2$ LL-like state remains an open and unsolved issue of great current interest.

  Surprisingly, subsequent experiments revealed similar ``two relaxation rates'' in $D=3$ correlated systems as well.  Specifically, simultaneous resistivity and Hall measurements in the classic Mott system 
V$_{2-y}$O$_{3}$~\cite{rosenbaum} revealed the following: in the lightly doped ($0<y<<1$) case, $(i)$ the dc 
resistivity, $\rho_{\text{dc}}(T)=\rho_{0}(y)+A(y)T^{1+\eta(y)}$, with $\eta(y)\leq 1.5$, while the Hall angle, 
cot$\theta_{\text{H}}(T)\simeq C_{1}+C_{2}T^{2}$, {\it independent} of $y$ for all $T>T_{N}(y)$, the N\'eel ordering temperature.  This is the first known example of a $D=3$ correlated metallic system exhibiting ``two'' relaxation rates, and similar behavior is also seen in nearly cubic CaRuO$_{3}$~\cite{laad:ruthenates} and YbRh$_{2}$Si$_{2}$~\cite{steglich:group}.  These observations show that
such novel features are not unique to $D=2$ systems, but generic to metallic states on the border of the Mott MIT.
It is also interesting~\cite{carter:honig1991} that disorder seems to be a very relevant perturbation in V$_{2-y}$O$_{3}$:
the resistivity is well accounted for by a variable-range hopping form, attesting to importance of disorder near the
Mott transition.
In multi-orbital CaRuO$_{3}$ and YbRh$_{2}$Si$_{2}$, orbital-selective physics~\cite{laad:ruthenates,werner:spinfreezing} generically 
leads to extinction of LFL metallicity via ``Kondo breakdown'' and onset of ``spin freezing'', wherein one would 
expect low-energy charge dynamics to be 
controlled by the (strong) ``intrinsic disorder'' scattering between the quasi-itinerant and effectively Mott 
localized components of the full one-particle spectral function (though, strictly speaking, consideration of 
YbRh$_{2}$Si$_{2}$ would require a multi-band periodic Anderson model) .  The actual Mott transition in 
V$_{2}$O$_{3}$ is by now also established to involve multi-orbital correlations and orbital-selective localization: 
in LDA+DMFT studies~\cite{laad:v2o3,held,georges:group}, the $e_{g}^{\pi}$ states remain Mott localized, while the 
$a_{1g}$ states remain bad-metallic in the bad-metal close to the MIT.  In the quantum paramagnetic state where the 
Mott transition occurs, one may view the $e_{g}^{\pi}$ states as providing an ``intrinsic disorder'', providing a 
strong scattering channel for the $a_{1g}$ carriers.  Thus, it seems that the anomalous two-relaxation times are 
linked to the breakdown of LFL metallicity arising from strong scattering processes involving either
intrinsic scattering channels or extrinsic disorder close to the MIT.

Motivated by these observations, we introduce a Hubbard-Falicov-Kimball model in standard notations
\begin{align}
H_{\HFKM}&=-t\sum_{\langle i,j\rangle,\sigma}(c_{i\sigma}^{\dag}c_{j\sigma}+\hc)\nonumber\\ 
&+\quad U\sum_{i}n_{i\uparrow}n_{i\downarrow} + U_{cd}\sum_{i,\sigma}n_{ic\sigma}n_{id}
\end{align}
as an \emph{effective} model that captures the interplay between itinerancy ($t$) and strong electronic correlations ($U$)
and intrinsic or extrinsic ($U_{cd}$) disorder scattering.  Qualitatively, (i) $U_{cd}$ can mimic an effectively 
Mott-localized band in an orbital-selective Mott transition (OSMT) scenario, or $(ii)$ $v_{i}=U_{cd}n_{id}$ can also be viewed as an extrinsic disorder potential experienced by the correlated $c$-fermions (in V$_{2-y}$O$_{3}$, one can regard this as disorder arising 
from a concentration $y$ of V-vacancies in the host system).  We solve $H_{\HFKM}$ using the dynamical mean-field theory (DMFT) with iterated perturbation theory (IPT) as the solver for the effective impurity problem~\cite{hbar:raja:ijmpb}. In our method, $U_{cd}$ is treated as site-diagonal disorder within coherent potential approximation (CPA)~\cite{laad2001} using a semi-circular band density of states for the $c$-electrons as an 
approximation to the actual $D=3$ system (it keeps the correct energy dependence near the band edges in $D=3$).
Within DMFT, it is long known that a correlated LFL metal for small $U_{cd}$ smoothly goes over to an incoherent 
bad metal without LFL quasiparticles as $U_{cd}$ increases.  Motivated by the fact that two-relaxation times seem to 
be linked to proximity to the (pure or selective) Mott transition, we focus on the evolution of the (magneto)-transport
when $U_{cd}$ is cranked up in the regime where $U/t=3.3$ is chosen to be close to the critical $(U/t)_{c}\simeq 3.4$
where a purely correlation-driven Mott transition~\cite{DMFT:RMP1996} obtains.  
Since the relevant DMFT formalism and the  associated equations have already been discussed in Ref.~\citenum{laad2001} in the related context of a binary-alloy disordered Hubbard model (the same IPT+CPA also solves the HFKM {\it exactly} in DMFT for the $c_{\sigma}$-fermions), we do not repeat them here.

\begin{figure}[!htp]
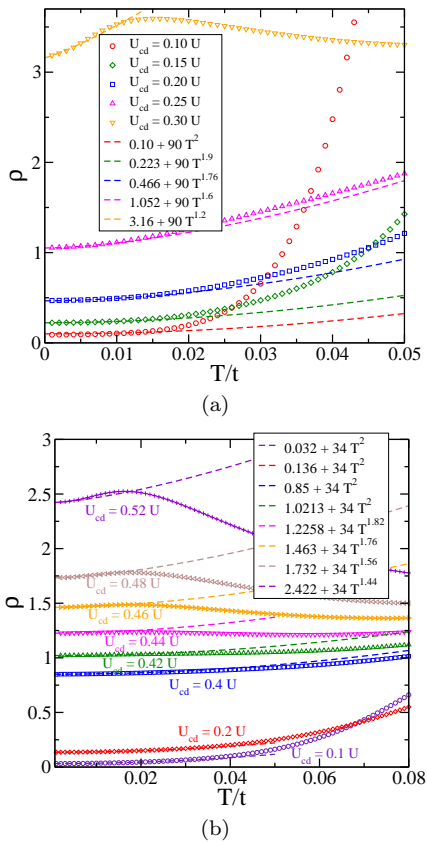

\subfigure[]{
\includegraphics[height=5cm,clip]{{01a_res_vs_T_U3.3}.eps}
}
\subfigure[]{
\includegraphics[height=5cm,clip]{{01b_res_vs_T_U2.0}.eps}
}
\caption{DC resistivity vs temperature plot at (a) $U=3.3 t$ and (b) $U=2.0 t$. Dashed lines are power-law fits at low $T$.}
\label{fig1:dc:res}
\end{figure}

   In Fig.~\ref{fig1:dc:res}, we exhibit the $\dc$ resistivity, $\rho_{\dc}(T,U_{cd})$ for different $U_{cd}$ and fixed $U/t=3.3$ as a function of $T$.  Several features stand out clearly: a correlated LFL up to small $U_{cd}<0.2U$, where
$\rho_{\dc}=\rho_{0}(U_{cd})+A(U_{cd})T^{2}$, smoothly evolves into an incoherent metal for $U_{cd}=0.2U$, where
we find $\rho_{\dc}(T)=\rho_{0}+A_{1}T^{\alpha}$, with $\alpha=1.76$.  It is very interesting that $\alpha$ seems to 
vary continuously with $U_{cd}$, ($\alpha=1.6$ for $U_{cd}=0.25U$, $\alpha=1.2$ for $U_{cd}=0.3U$), {\it and} the 
fact that $\rho_{dc}$ remains bad-metallic at intermediate-to-low $T$, crossing over to an insulator-like form at 
high $T$ for $U_{cd}=0.3U$.  Repeating the calculations for smaller $U/t=2.0$, we find that while qualitatively similar features obtain, $\rho_{dc}(T\rightarrow 0)$ rises to much higher values when $U_{cd}$ is cranked up.  This 
testifies to the increasing relevance of the strong scattering (from localized channels or extrinsic disorder as above)
when $U/t$ is in the weak-to-intermediate coupling regime.  Since transport properties in DMFT do not involve vertex
corrections in the Bethe-Salpeter equations for the conductivities, these features must be tied to loss of the LFL
quasiparticle pole structure in the DMFT one-electron propagator, which is now the sole input to the renormalized 
bubble diagram for the conductivities~\cite{DMFT:RMP1996}.  

\begin{figure}[!htp]
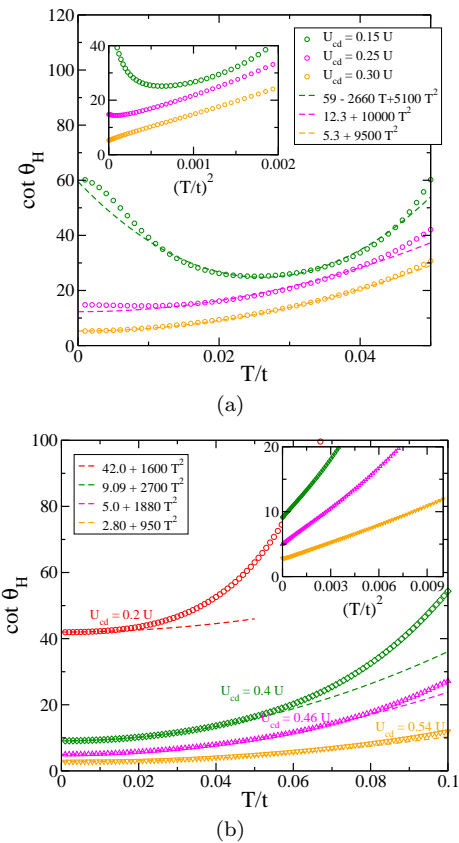

\subfigure[]{
\includegraphics[height=5cm,clip]{{02a_cotthetaH_vs_T_U3.3}.eps}
}
\subfigure[]{
\includegraphics[height=5cm,clip]{{02b_cotthetaH_vs_T_U2.0}.eps}
}
\caption{Cotangent of Hall angle ($\theta_H$) vs temperature plot at (a) $U=3.3 t$ and (b) $U=2.0 t$. Dashed lines are power-law fits at low $T$. Insets show the same against $T^2$.}
\label{fig2:cotthetaH:vs:T}
\end{figure}

\begin{figure}[!htp]
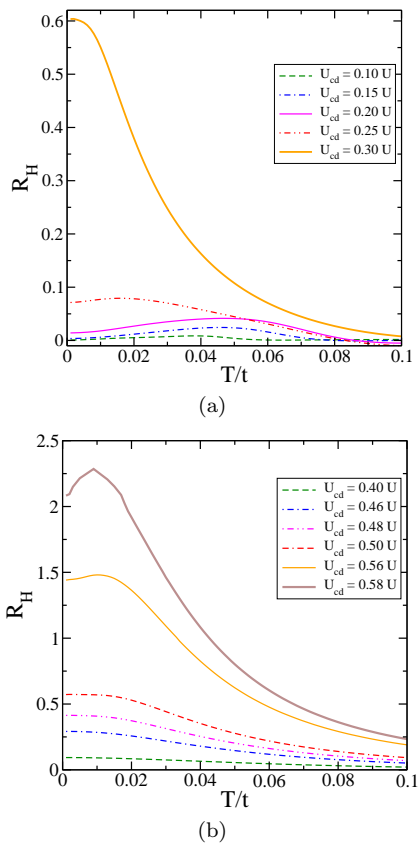

\subfigure[]{
\includegraphics[height=5cm,clip]{{03a_RH_vs_T_U3.3}.eps}
}
\subfigure[]{
\includegraphics[height=5cm,clip]{{03b_RH_vs_T_U2.0}.eps}
}
\caption{Hall coefficient ($R_H$) vs temperature plot at (a) $U=3.3 t$ and (b) $U=2.0 t$. Dashed lines are power-law fits at low $T$.}
\label{fig3:RH:vs:T}
\end{figure}

   Upon evaluating the off-diagonal conductivity (to first order in the vector potential ${\bf A}$ as done before~\cite{lange:kotliar1999}) for $H_{\HFKM}$ in DMFT, we have computed the Hall constant ($R_{\text{H}}$) and Hall angle (cot$\theta_{\text{H}}$) for the 
same parameter values as above.  Even more remarkably, we find (see Fig.~\ref{fig2:cotthetaH:vs:T}) that cot$\theta_{\text{H}}\simeq C_{1}+C_{2}T^{2}$, up to $T/t=0.05$ for both $U_{cd}/U=$ 0.25, 0.3, while $R_{\text{H}}$ exhibits a strong $T$-dependence right down to 
 the lowest $T$ (see Fig.~\ref{fig3:RH:vs:T}).  This is the {\it same} parameter regime where $\rho_{dc}(T)$ exhibits non-LFL
$T$-dependence, with a $U_{cd}$-dependent exponent $1.0 <\alpha < 2.0$.  Thus, our DMFT results directly reveal 
two-relaxation rates, and it is indeed notable that cot$\theta_{\text{H}}\simeq C_{1}+C_{2}T^{2}$ continues to hold over a 
wide $T$ range, even as the exponent $\alpha$ continuously varies between $1.0$ and $2.0$.  Our results are completely
consistent with data for V$_{2-y}$O$_{3}$~\cite{rosenbaum} in all respects: $(i)$ specifically, $\rho_{dc}(T)\simeq
\rho_{0}(y)+AT^{\alpha}$ with $1.3\leq\alpha\leq 1.5$ in data agrees well with our estimate $1.2\leq 1.76$ in the 
non-LFL regime of $H_{\HFKM}$, $(ii)$ cot$\theta_{\text{H}}(T)=C_{1}+C_{2}T^{2}$ up to $T\simeq 500$~K upon choosing 
$t=1.0$~eV in the model, again in nice accord with data.  Moreover, cot$\theta_{\text{H}}$ also exhibits an upward curvature 
at very low $T$ in DMFT, again in complete accord with data.  $(iii)$ Concomitantly, $R_{\text{H}}(T)$ exhibits a strong 
$T$-dependence, increasing with decreasing $T$ before peaking at very low $T$ {\it before} AF order occurs at 
$T\simeq 10$~K. 

   Thus, (magneto)transport responses in $H_{\HFKM}$ within DMFT exhibit comprehensive qualitative accord with the 
complete set of data for V$_{2-y}$O$_{3}$.  In particular, our results now strongly support the notion that emergence
of two relaxation rates, or that the different decay rates for longitudinal and Hall currents, is a direct 
consequence of breakdown of the LFL metal by strong scattering.  As discussed above, the FK term in Eq.(1) can mimic 
either ``intrinsic'' scattering coming from selectively Mott localized states in a multi-band system, or arising from 
strong ``disorder'' scattering.  
Indeed, singular behavior of the $\gamma$-coefficient of the specific heat~\cite{poteryaev2015} has also been recently found in a binary-alloy Hubbard model: we emphasize that this is isomorphic to our HFKM, since the FK term can also be interpreted as a “binary alloy disorder” term
in the Hubbard model. Thus, a similar divergence of $\gamma(T)=C_{\text{el}}(T)/T$ will also
appear in our HFKM within DMFT. This generic effect of a strong intrinsic (or extrin-
sic, of the binary-alloy disorder type) will be generally relevant close to a correlation-driven MIT.
This is because correlations have already drastically renormalized the band energy scale to a much 
lower value associated with collective Kondo screening induced ``heavy'' LFL.  In such a situation, even modest 
disorder will appear ``strong'', since the relevant scale that sets the relevance of disorder is now $(U_{cd}n_{id}/k_{B}T_{\text{coh}})$ (with coherence temperature $T_{\text{coh}}$ being small near the Mott transition) rather than $(U_{cd}n_{id}/W)$, with $W$ the 
free bandwidth for $U=0$.  

   These observations call for a deeper understanding in terms of basic microscopic responses involving interplay 
between the FK term and Mottness.  Since DMFT is a self-consistently embedded impurity problem, and the anomalies 
are seen in the strongly correlated metallic state, we choose to tease out the deeper underlying reasons by 
analyzing the ``impurity'' model.  In analogy with DMFT studies for the Hubbard model, the appropriate impurity model
is the Wolff model of a correlated $d$-impurity coupled to a bath of ``conduction'' electrons, as well as to a 
localized scattering potential ($U_{cd}$).  We find it most convenient to bosonize this impurity model by generalizing
earlier attempts~\cite{gmzhang}.  Such an analysis has the potential to bare the asymptotic separation of spin- and charge
modes, facilitating DMFT observation of two relaxation rates.  The Wolff impurity model including the FK coupling reads
\begin{align}
H_{\text{W}}&=\sum_{k,\sigma}\epsilon_{k}c_{k\sigma}^{\dag}c_{k\sigma} + Un_{0,d,\uparrow}n_{0,d,\downarrow} + U_{cd}\sum_{\sigma} n_{0,c,\sigma}n_{0,d}\nonumber\\
&\quad-\mu\sum_{\sigma}n_{0,c,\sigma}\,.
\end{align}
The Wolff model for $U_{cd}=0$ is bosonized as usual on a $(1+1)$ $D$ half-line and the result is a set of two 
independent gaussian models for {\it bosonic} spin and charge fluctuation modes emanating from the ``impurity'' (site $0$) in 
each radial direction.  The bosonized Hamiltonian is $H=H_{c}+H_{s}$ where

\begin{align}
H_{c}&=\frac{v_F}{2}\int dx\, \bigg[\Pi_{c}^{2}(x)+(\partial_{x}\phi_{c})^{2}\bigg] + \frac{(\mu+Un_{0})}{\sqrt{2}\pi}[\partial_{x}\phi_{c}(0)]\nonumber\\ 
&\quad+ \frac{U}{8\pi^{2}}[\partial_{x}\phi_{c}(0)]^{2}\,,\\
H_{s}&=\frac{v_F}{2}\int dx\, \bigg[\Pi_{s}^{2}(x)+(\partial_{x}\phi_{s})^{2}\bigg] - \frac{U}{8\pi^{2}}[\partial_{x}\phi_{s}(0)]^{2}
\end{align}
for a non-magnetic ground state, with $n_{0}$ corresponding to the occupation of non-interacting $c$-fermions.  The spin 
and charge bosonic fields on the impurity are $\rho_{s}(0)=(\partial_{x}\phi_{s})/\sqrt{2\pi}$ and $\rho_{c}(0)=(\partial_{x}\phi_{c})/\sqrt{2\pi}$.  $v_{F}$ is the (common) Fermi velocity when $U_{cd}=0$.  The FK term couples solely to the charge bosons, 
but with a subtlety which will result in interesting
anomalies.  Viewed at the basic scattering level, $U_{cd}\sum_{i}n_{ic}n_{id}$ acts as a strong scattering potential.  Since
$[n_{id},H]=0\, \forall i$ in the HFKM, $n_{id}=0,1$ and, viewed by a propagating $c$-fermion, $v_{i}=U_{cd}n_{id}$ now 
represents a ``suddenly switched on'' scattering potential that successively switches suddenly between $0$ and $U_{cd}$.  In 
the local impurity problem, this is thus precisely the famed ``X-ray edge'' (XRE) problem ($U_{cd}$ is the precise analog of the
suddenly switched-on ``core-hole'' potential) in the {\it charge} channel: the spin channel is left unaffected.  
In their seminal works, Anderson~\cite{anderson:xre1} and Nozieres {\it {\it et al}.}~\cite{nozieres:xre2}
found vanishing overlap between ground states with $U_{cd}=0$ and $U_{cd}\neq 0$, and emergence of infra-red branch-cut features
in one- and two-particle propagators.  Fortunately, the XRE problem is also readily amenable~\cite{schotte} to bosonization:
the crucial effect of $U_{cd}$ is to induce a ``shift'' in the charge bosons.  Explicitly, expanding the charge-bosonic field in Fourier components, $\phi_{c}(x)=\sum_{k}(\sqrt{2|k|}^{-1}(a_{k}e^{ikx}+a_{k}^{\dag}e^{-ikx})e^{-\alpha|k|/2}$ and
$\Pi_{c}(x)=-i\sum_{k}\sqrt{|k|/2}(a_{k}e^{ikx}-a_{k}^{\dag}e^{-ikx})e^{-\alpha|k|/2}$, one gets

\begin{align}
H_{c}&=\sum_{k>0}\omega_{k}a_{k}^{\dag}a_{k} + \frac{iv\sqrt{2\rho}}{\sqrt{N}}\sum_{k>0}(a_{k}-a_{-k})\nonumber\\
&\quad-\frac{U\rho}{2}\sum_{k,k'>0}(a_{k}-a_{-k})(a_{k'}-a_{-k'})
\end{align}
where $a_{k}^{\dag}=a_{-k}$, we have set $\mu=U(1-n_{0})$ using Luttinger's theorem, and work within the restricted Hilbert 
space where the $b_{k},b_{k'}^{\dag}$ satisfy Bose commutation relations.  $H_{c}$ now corresponds to a shifted oscillator, 
and the effect of $U_{cd}$ is to generate an unrenormalizable $s$-wave phase shift, $\delta=$tan$^{-1}(U_{cd}/2z_{\FL}t)$, with
$z_{\FL}$ the quasiparticle renormalization in the disorder free (pure Hubbard) model.

  $H_{c}$ is now a ``small polaron'' model with coupling between bosons at different $k$ along different directions from the 
impurity, where the ``polarons'' are now associated with low-energy particle-hole modes.  Since it is quadratic 
in the bosons, it can be readily diagonalized as follows: $(i)$ the first ``small polaron'' like term is transformed away by a 
Lang-Firsov unitary transformation, resulting in a shifted ``oscillator'' form~\cite{schotte}, resulting in 
$H_{c}=\sum_{k>0}(k/\rho)(a_{k}^{\dag}+U_{cd}/\sqrt{kN})(a_{k}+U_{cd}/\sqrt{kN})$ and $(ii)$ the last term, 
quadratic in bosons but with mixing terms such as $a_{k}^{\dag}a_{k'}^{\dag}$ and $a_{k'}a_{k}$, simply rotates $H_{c}$ to
a new quadratic Hamiltonian in bosons.  However, the crucial effect of the last step is to change the velocity of the 
charge bosons (the sole but crucial effect of $U_{cd}$ is, via action of the term $(a_{k}-a_{-k})=(a_{k}+a_{k}^{\dag})$,
 to ``distort'' the Luttinger Fermi sea for the charge, but {\it not} for spin): thus, this results in different velocities
for charge and spin modes, with $v_{c}>v_{s}$.   
The charge-bosonic propagator will acquire an anomalous dimension, while the spin-fluctuation propagator will retain its ``Fermi 
liquid'' like character.  Put another way, the charge fluctuation propagator now has a branch-cut, but the spin fluctuation 
propagator retains its infra-red pole structure.  Upon refermionization of $H_{s}$ (note that this {\it cannot} be done for 
$H_{c}$ in view of the orthogonality catastrophe), we find that the spin excitations are expressible as {\it fermions}: 
following Anderson, one may call them ``spinons''.  Remarkably, this bears close similarity to Anderson's 
hidden-FL~\cite{anderson:hiddenFL}, and the above can be viewed as a high-dimensional spin-charge separation.

  An external electric field accelerates charge, leading to a spinon backflow and induces scattering between spin and charge. In $D=3$, one expects
that scattering off local, dynamical spin fluctuations will lead to the $\dc$ resistivity $\rho_{\dc}(T)\simeq T^{D/2}=T^{3/2}$.  
However, an external magnetic field will couple solely to the spin modes (equivalently spin-fermions or ``spinons'' as above), 
leading to a Hall relaxation rate entirely determined by ``spinon-spinon'' scattering, giving 
cot$\theta_{\text{H}}\simeq \tau_{\text{H}}^{-1}=C_{2}T^{2}$.  In presence additional strong scattering due to either an intrinsically  
localized electronic ($U_{cd}$) or disorder channel, there will generically be a term $\tau_{\text{H}}^{-1}\simeq C_{1}$ in addition 
to the above, yielding cot$\theta_{\text{H}}\simeq C_{1}+C_{2}T^{2}$.  On the other hand, since the charge fluctuations are directly 
affected thereby, the resulting modification of scattering processes involving charge and spin modes can lead to deviation from
the $\rho_{\dc}(T)\simeq T^{3/2}$ in addition to contributing a residual $\rho_{0}$ term.

   The finding of two relaxation rates for decay of longitudinal and Hall currents can now be rationalized by observing that 
these are consequences of breakdown of LFL concepts in the barely (bad) metallic state close to the Mott transition.  Extinction
of LFL quasiparticles is associated with a ``lattice'' orthogonality catastrophe, which now occurs due to either $U_{cd}$~\cite{georges:fkm} or strong disorder~\cite{altshuler} in a metal already close to a correlation-driven Mott transition.
In this context, it is interesting to observe that the hidden-FL~\cite{anderson:hiddenFL} also involves a related X-ray-edge mechanism 
(at $U=\infty$) for destruction of LFL theory.  In our case, given finite $U\simeq W$ (the one-electron bandwidth), additional 
intrinsic ($U_{cd}$) or extrinsic (disorder) scattering channels are necessary to generate such breakdown of LFL theory.    
 Turning to $D=2$, observation of similar features in near-optimally doped cuprates will require appeal to cluster extensions
of DMFT (which, among other things, cannot access dynamical effects of non-local spatial correlations near a Mott transition).
However, our use of DMFT is known to be a reliable approximation for $D=3$.

   To summarize, we have investigated emergence of two relaxation rates in correlated metals close to the Mott transition in
$D=3$.  We find that this unique feature is tied to loss of LFL metallicity in symmetry-unbroken metallic states proximate to
Mott transition(s): this can arise from strong scattering processes either stemming from intrinsic, (selectively-Mott) localized
electrons, or from disorder which is generically relevant near a MIT.  It is thus not specific to $D=2$.  Surprisingly, 
comparison with data~\cite{rosenbaum} for V$_{2-y}$O$_{3}$ reveals very good qualitative accord with all unusual features:
$(i)$ $\rho_{\dc}\simeq \rho_{0}(y)+AT^{\alpha(y)}$ with $1.2\leq\alpha\leq 1.6$, $(ii)$ a strong $T$-dependent Hall constant,
peaking at low $T$, and $(iii)$ much more disorder-independent behavior of cot$\theta_{\text{H}}(T)\simeq C_{1}+C_{2}T^{2}$.

\begin{acknowledgments}
We are thankful to the DAE, Govt. of India for the financial support.
\end{acknowledgments}

\bibliographystyle{apsrev4-1}

\end{document}